\documentclass[aps,prb,reprint]{revtex4-1}

\usepackage[utf8]{inputenc}
\usepackage{xcolor}
\usepackage{makecell}
\usepackage{graphicx}
\usepackage{float}
\usepackage{amsmath}

\newcommand{\X}{\mathbf{X}}

\newcommand{\FM}{\mathbf{F}^\text{M}}
\newcommand{\FG}{\mathbf{F}^\text{G}}
\newcommand{\FD}{\mathbf{F}^\text{D}}
\newcommand{\FRes}{\mathbf{F}^\text{R}}
\newcommand{\Fext}{\mathbf{F}^\text{ext}}

\newcommand{\Ms}{M_\text{s}}
\newcommand{\lex}{l_\text{ex}}
\newcommand{\HOe}{\mathbf{H}_\text{Oe}}

\newcommand{\kex}{k^\text{ex}}
\newcommand{\kOe}{\kappa^\text{Oe}}
\newcommand{\kms}{k^\text{ms}}
\newcommand{\Wex}{W^\text{ex}}
\newcommand{\WOe}{W^\text{Oe}}
\newcommand{\Wms}{W^\text{ms}}

\newcommand{\Cp}{C^+}

\newcommand{\Cm}{C^-}

\newcommand{\titre}{Quantitative and realistic description of the magnetic potential energy of spin-torque vortex oscillators}

\begin{document}

\title{\titre}
\author{Simon DE WERGIFOSSE}
\email{simon.dewergifosse@uclouvain.be}
\affiliation{Institute of Condensed Matter and Nanosciences, Universit\'{e} catholique de Louvain, Place Croix du Sud 1, 1348 Louvain-la-Neuve, Belgium}
\author{Chloé CHOPIN}
\affiliation{Institute of Condensed Matter and Nanosciences, Universit\'{e} catholique de Louvain, Place Croix du Sud 1, 1348 Louvain-la-Neuve, Belgium}
\author{Flavio ABREU ARAUJO}
\affiliation{Institute of Condensed Matter and Nanosciences, Universit\'{e} catholique de Louvain, Place Croix du Sud 1, 1348 Louvain-la-Neuve, Belgium}

\begin{abstract}
Understanding the dynamics of magnetic vortices has emerged as an important challenge regarding the recent development of spin-torque vortex oscillators.  Either micromagnetic simulations or the analytical Thiele equation approach are typically used to study such systems theoretically. This work focuses on the precise description of the restoring forces exerted on the vortex when it is displaced from equilibrium. In particular, the stiffness parameters related to a modification of the magnetic potential energy terms are investigated. A method is proposed to extract exchange, magnetostatic and Zeeman stiffness expressions from micromagnetic simulations. These expressions are then compared to state-of-the-art analytical derivations. Furthermore, it is shown that the stiffness parameters depend not only on the vortex core position but also on the injected current density. This phenomenon is not predicted by commonly used analytical ansätze. We show that these findings result from a deformation of the theoretical magnetic texture caused by the current induced Amp\`ere-Oersted field.
\end{abstract}

\maketitle

\section{Introduction}
Magnetic vortices are one of the encountered magnetic ground states in soft ferromagnets of reduced dimensions. These non-uniform topologies are characterized by curling in-plane spins swirling around a small area called the vortex core, where the magnetization points out-of-plane. Such distribution results from a favorable trade-off between exchange and magnetostatic energies, arising for a wide range of nanodot aspect ratios.\cite{metlov2002stability} Two main topological parameters are typically used to describe a vortex,\cite{guslienko2006low} namely the chirality $C$ (also called circulation or helicity) and the polarity $P$. On the one hand, the chirality indicates whether the in-plane magnetization curls clockwise ($C=-1$) or counterclockwise ($C=+1$). On the other hand, the polarity informs on the direction of the out-of-plane vortex core profile, which either points upwards ($P=+1$) or downwards ($P=-1$). Due to their great stability,\cite{hertel2007ultrafast,guslienko2008magnetic,hrkac2015magnetic} vortices have rapidly gained a lot of attention for prospective applications, notably when combined with spintronics.

Indeed, concomitantly to the advances in vortex dynamics understanding were investigated the first spin-torque nano-oscillators.\cite{katine2000current,kiselev2003microwave,slavin2009nonlinear,zeng2013spin} This type of device, based on a magnetic tunnel junction structure, allows to inject a spin-polarized current into a free ferromagnetic layer. As a spin-transfer torque is exerted on its magnetization by conservation of angular momentum, it allows to act on the direction of its mean magnetic moment. Based on magnetoresistance measurements, an alternating signal may thus be retrieved when injecting a constant current into these cylindrical heterostructures. This makes it one of the smallest dc to ac converters. As noted above, it may happen that, for some geometries, the free layer presents a vortex as its magnetic ground state. This led to the development of what are called spin-torque vortex oscillators (STVOs).\cite{pribiag2007magnetic,sluka2015spin} In the absence of any external excitation, the vortex core is located at the center of the nanodot. However, if a sufficient input current is applied to the heterostructure and providing the appropriate vortex polarity,\cite{araujo2022ampere} steady-state oscillations of magnetization may occur. These oscillations, usually in the hundreds of MHz range, are caused by a shift of the moving vortex core towards a non-zero orbit of precession. For very large excitations, the off-centered vortex core may even be expelled from the nanodot which can lead to damping, after the nucleation of a vortex of opposite polarity,\cite{araujo2022ampere,yamada2007electrical,guslienko2008dynamic,khvalkovskiy2009vortex,gaididei2010magnetic} or more exotic magnetic states.\cite{jenkins2016spin,wittrock2021beyond} Besides good stability, STVOs present many advantages,\cite{pribiag2007magnetic} {\it e.g.}, low noise sensibility, no external field required, narrow bandwidth and wide frequency tunability, which make them potentially appealing systems for radiofrequency\cite{jenkins2016spin} or artificial intelligence applications. \cite{torrejon2017neuromorphic, Romera2018}

As STVOs started to gain interest, theoretical studies on vortex dynamics became crucial to capture and predict device properties such as the frequency or emitted power. Micromagnetic simulations and the so-called Thiele equation\cite{thiele1973steady,thiele1974applications} framework are typically used to examine such systems. Among all aspects of vortex dynamics, the description of restoring forces associated to a displacement of the vortex core from its equilibrium position has been a particularly rich research topic. Stiffness parameters, associated to each magnetic potential energy terms, are commonly defined to characterize the restoring force, as it would be done for classical springs. These spring-like parameters have been extensively studied analytically.\cite{guslienko2001field,guslienko2006low,gaididei2010magnetic,araujo2022ampere} However very few groups have retrieved those from micromagnetic simulations.\cite{buchanan2006magnetic,choi2008understanding,fried2016localized} In this work, we will propose a method to extract expressions of the vortex stiffness, valid over its dynamic range, from simulations. We will then compare them with analytical derivations from the literature. Lastly, we will show the impact of the Amp\`ere-Oersted field on the magnetic texture.
\section{Methods}
\begin{figure}[ht]
    \centering
    \includegraphics{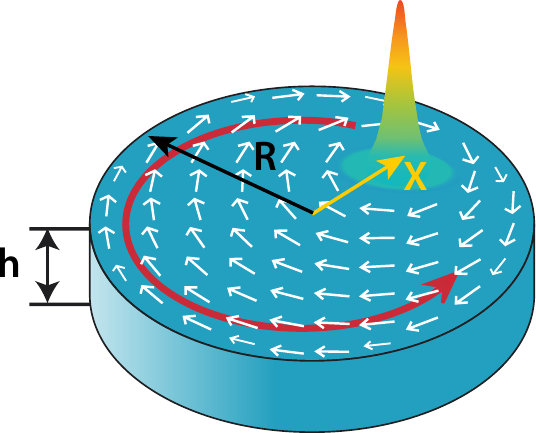}
    \caption{Schematic of the free layer of radius $R$ and thickness $h$ of the magnetic tunnel junction under study. The in-plane components of the vortex magnetization field $\bf{m}$ are represented by the white arrows. The precessional gyration of the vortex core, located at $\bf{X}$, is depicted by the red arrow. The color gradient represents the magnitude of the out-of-plane magnetization component $m_z$.}
    \label{fig:mtj}
\end{figure}

 The Thiele equation approach (TEA) is usually used to describe analytically the dynamics of magnetic vortices confined in the free layer of magnetic tunnel junctions (see Fig.~\ref{fig:mtj}), as first proposed by Huber\cite{huber1982dynamics} after works on bubble materials by Malozemoff \& Slonczewski.\cite{malozemoff1979magnetic} This theoretical framework allows to predict the vortex core in-plane position $\X=(X,Y)$ by looking at the sum of the forces acting on it. The core is thus seen as a quasi-particle representative of any modification of the global magnetic distribution. At equilibrium, the most general way to express this system \cite{guslienko2006low} is given as
\begin{equation}
    \FM + \FG + \FD + \FRes + \Fext = \mathbf{0},
    \label{eq:TEA}
\end{equation}
where $\FM$ is an inertial mass term, $\FG$ is the gyrotropic force related to the dominant excitation mode of the vortex at low frequency, $\FD$ is a Gilbert dissipation term, $\FRes$ are the restoring forces and $\Fext$ represents any additional external forces. For an isolated STVO, these external forces are mainly related to spin-transfer torques.\cite{slonczewski1996current,li2003magnetization}

 Any displacement of the vortex core from its original ground state is associated to a modification of the magnetic potential energy $W$. By analogy to simple harmonic motions, one can define a stiffness $k$ related to the system. Keeping all generality, this spring-like parameter is not constant but depends on the degree of deformation, {\it i.e.}, the vortex core position here. Its value allows to calculate the restoring forces appearing in Eq.~(\ref{eq:TEA}) as
\begin{equation}
    \FRes = -\frac{\partial W}{\partial \X} = -k\X.
    \label{eq:Fr}
\end{equation}

Neglecting magnetocrystalline anisotropy and any external magnetic field, the magnetic potential energy $W$ is composed of three terms, namely the exchange $\Wex$, magnetostatic $\Wms$ and Zeeman $W^\text{Z}$ ($=\WOe$) energies, the latter being thus only associated to the current induced Ampère-Oersted field (AOF). Based on these hypotheses, the total energy $W$ can be calculated as \cite{guslienko2001magnetization,gaididei2010magnetic}
\begin{equation}
    W = \int_V \left(A(\nabla \mathbf{m})^2-\frac{1}{2}\mathbf{M}\cdot\mathbf{H}^\text{ms}-\mathbf{M}\cdot\mathbf{H^\text{Oe}}\right) dV,
    \label{eq:W}
\end{equation}
where $V$ is the volume of the magnetic dot, $\mathbf{m}=\mathbf{M}/\Ms$ is the normalized magnetization, with $\Ms$ the saturation magnetization, $A$ is the exchange stiffness coefficient and $\mathbf{H}^\text{ms}$ and $\mathbf{H}^\text{Oe}$ are the magnetostatic and AO fields, respectively. This paper is written using cgs units. A conversion table between SI and cgs systems in magnetism is available in ref.\cite{blundell2001magnetism}, if necessary.

To proceed to further calculations from Eq.~(\ref{eq:W}), a mathematical description of the magnetization distribution for a vortex state in circular nanopillars of radius $R$ is required. In this respect, the so-called two vortex ansatz (TVA), developed by Guslienko {\it et al.},\cite{guslienko2001evolution,guslienko2002eigenfrequencies} has been widely used to study the dynamics of vortices excited by out-of-plane spin-polarized currents.\cite{ivanov2007excitation,khvalkovskiy2009vortex,belanovsky2012phase,dussaux2012field,guslienko2014nonlinear,araujo2022ampere} This theoretical model relies on the combination of two off-centered rigid vortices with cores located at  $\X=(\rho,\varphi)$ and $\X_\text{I}=(R^2/\rho,\varphi)$, in polar coordinates. Contrary to former descriptions, it presents the advantage of meeting Dirichlet boundary conditions, {\it i.e.}, no net magnetic charge at the surface edge of the dot. Following TVA, the magnetization at any point $\mathbf{r}$ in the nanodot can be characterized by a planar angle $\phi$ and core profile angle $\Theta$. Those are expressed as
\begin{equation}
    \begin{cases}
        \phi = \arg(\mathbf{r}-\mathbf{X})+\arg(\mathbf{r}-\mathbf{X}_\text{I})-\varphi+C\frac{\pi}{2}\\
        \Theta = \arccos(Pf_z(||\mathbf{r}-\X||)) 
    \end{cases}
    \label{eq:angles}
\end{equation}
where $f_z$ is a function describing the vortex core profile. Various propositions of bell-shaped curves $f_z$ have been reported in previous works.\cite{gaididei2010magnetic,usov1993magnetization,hoellinger2003statics} Straightforwardly, one can express the three spatial components of the magnetization $\mathbf{m}$ using these angles as 
\begin{equation}
    \mathbf{m} = \begin{bmatrix}m_x\\m_y\\m_z \end{bmatrix}=\begin{bmatrix} \cos \phi \sin \Theta \\ \sin \phi \sin \Theta \\ \cos\Theta \end{bmatrix}.
    \label{eq:m}
\end{equation}
Using TVA and Eq.~(\ref{eq:W}), developments have been undertaken in the past two decades to obtain expressions of the stiffness parameters $\kex$, $\kms$ and $k^\text{Oe}$, related to each energy component, as a function of the reduced orbit radius $s=||\X||/R$.

Gaididei {\it et al.}\cite{gaididei2010magnetic} obtained the following expression for the exchange contribution, based on developments from Guslienko {\it et al.},\cite{guslienko2001field}
\begin{equation}
    \kex = 8 \pi^2 h \Ms^2 \left(\frac{\lex}{R}\right)^2 \frac{1}{1-s^2},
    \label{eq:kexT}
\end{equation}
where $\lex=\sqrt{A/(2\pi\Ms^2)}$ is the exchange length of the material and $h$ is the thickness of the free layer. Concerning the magnetostatic term, we recently solved\cite{araujo2022ampere} the expression of the energy $\Wms$ proposed by Gaididei {\it et al.},\cite{gaididei2010magnetic} which led to
\begin{equation}
    \kms_\xi = \frac{8\Ms^2 h^2}{R}\textstyle \Lambda_{0,\xi}\left(1+a_\xi s^2+b_\xi s^4 + c_\xi s^6\right)
    \label{eq:kmsT}
\end{equation}
where $\Lambda_{0,\xi}$, $a_\xi$, $b_\xi$ and $c_\xi$ are coefficients that can be calculated numerically for each value of the nanodot aspect ratio $\xi=h/(2R)$. Finally, we developed in the same study\cite{araujo2022ampere} an expression for the contribution associated to the AOF, $\kOe=k^\text{Oe}/(CJ)$, with $J$ being the current density imposed. It is expressed as
\begin{equation}
    \kOe = \frac{8\pi^2}{75}\Ms R h \left(1-\frac{4}{7}s^2 - \frac{1}{7} s^4 - \frac{16}{231} s^6 - \frac{125}{3003}s^8 \right).
    \label{eq:kOeT}
\end{equation}

In addition to these analytical works, micromagnetic simulations (MMS) have been extensively used to investigate STVO dynamics. Those rely on solving the Landau-Lifshiftz-Gilbert-Slonczewski\cite{landau1935theory,gilbert2004phenomenological,slonczewski1996current} equation to predict numerically the properties of magnetic systems of reduced dimensions. Despite the fact that TEA derives from the LLGS formalism, the latter relies on the use of an effective magnetic field rather than spring-like constants. Consequently, there is no direct access to stiffness parameters from micromagnetism. However, as the orbit radius $s$ and the energy components are retrievable from most solvers, these parameters can still be calculated by following the procedure described herebelow. As a first step, the energy must be fitted to a chosen expression, as a function of the vortex core position. Then, these functions must be derived according to Eq.~(\ref{eq:Fr}) to find the corresponding stiffness parameter. This strategy has already been successfully applied in previous studies,\cite{buchanan2006magnetic,choi2008understanding,fried2016localized} although for a limited $s$-range.

\begin{figure}[ht]
    \centering
    \includegraphics[width=0.45\textwidth]{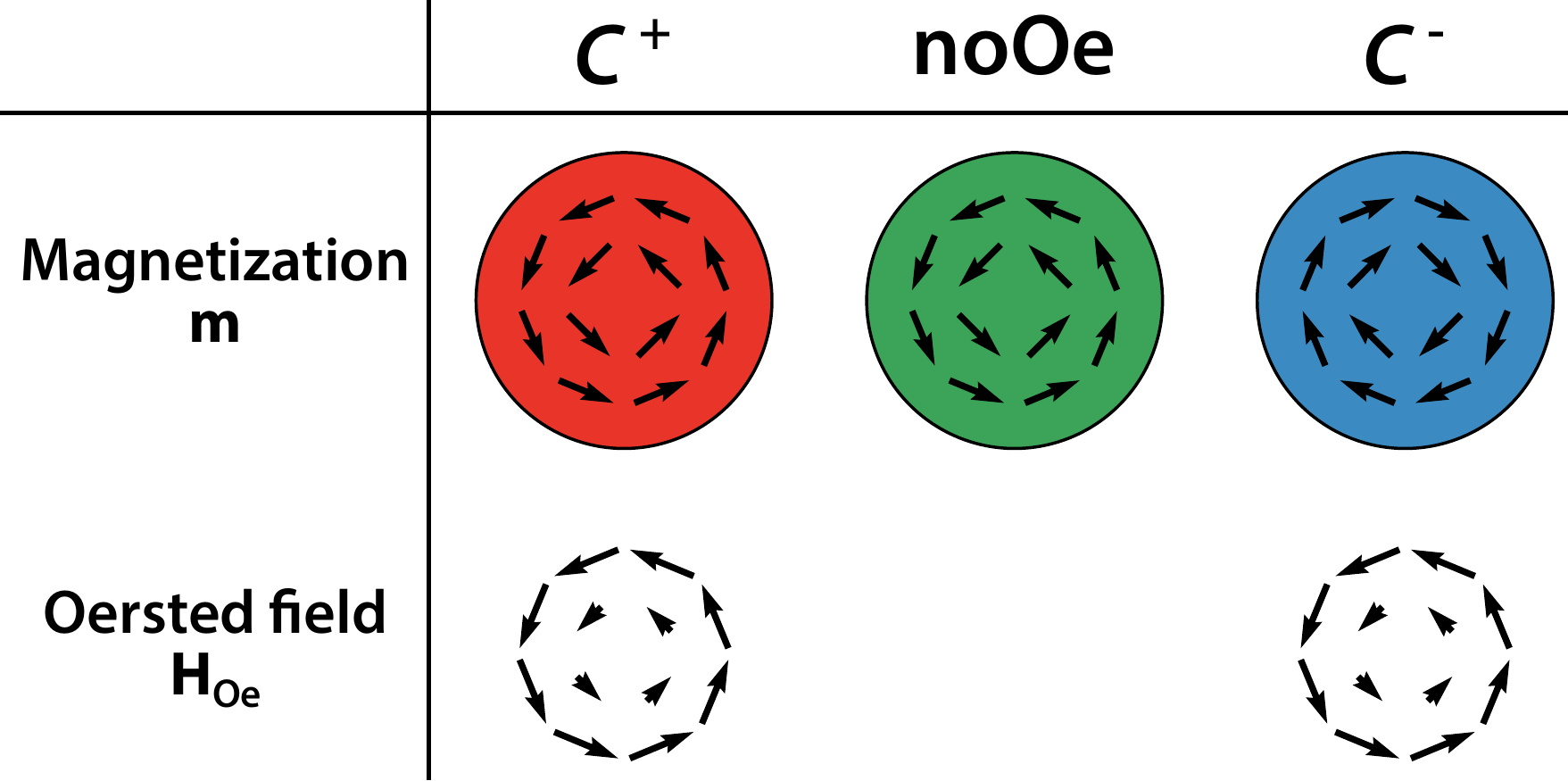}
    \caption{Relative orientation between the vortex in-plane magnetization $\bf{m}$ and the Amp\`ere-Oersted field (AOF) $\HOe$ in the three configurations investigated ({\it i.e.}, $C^+$, noOe and $C^-$, in red, green and blue, respectively). Considering an out-of-plane positive current, the curling magnetization is parallel, (resp. anti-parallel) to the AOF if the chirality is positive (resp. negative). For noOe, the AOF is not taken into account.}
    \label{fig:configs}
\end{figure}

In our case, the GPU-based solver MuMax3 is used to perform micromagnetic simulations.\cite{vansteenkiste2014design} A magnetic tunnel junction presenting a free layer of radius $R=100$~nm and thickness $h=10$~nm is studied. The latter is discretized into cells of dimensions $2.5\times2.5\times5$~nm$^3$. Typical material parameters for permalloy are used.\cite{cowburn2000property,vaz2008magnetism,guimaraes2009principles} The saturation magnetization and exchange stiffness coefficient are $\Ms=800$~emu/cm$^3$ and $A=1.07\cdot 10^{-6}$~erg/cm, respectively. The Gilbert damping constant $\alpha_\text{G}$ is fixed at 0.01 and the spin-current polarization $p_J$ at 0.2. A polarizer presenting a perpendicular direction of magnetization $\mathbf{p}=(p_x,p_y,p_z)=(0,0,1)$ is used. MuMax3 includes both Slonczewski\cite{slonczewski1996current} and Zhang-Li\cite{zhang2004roles} spin-transfer torques. Nonetheless, given the out-of-plane current configuration, the field-like torque has a limited influence on the dynamics as the vortex profile is uniform along the free layer thickness for thin magnetic dots.\cite{guslienko2022magnetic} In light of these considerations, we have chosen to simplify the analysis by assuming an adiabatic situation, where the degree of non-adiabaticity is fixed at zero. The vortex polarity is chosen to be $P=-1$. The influence of the temperature is not taken into account in this study. Current densities of $2,~4$ and $6 \cdot 10^6$~A/cm$^2$ are imposed into the oscillator, in the positive $z$-direction which allows to respect one of the two necessary conditions for steady-state precessions\cite{araujo2022ampere} ({\it i.e.}, $JPp_z<0$). The second criterion is to inject a current exceeding the first critical current density $J_\text{c1}$. When $J<J_\text{c1}$, simulations are initiated with a vortex core translated to $s=0.9$ which damps back to the center of the dot, {\it i.e.}, $s=0$. For $J>J_\text{c1}$ however, two simulations are required to explore the whole range of $s$ for a given current, as the steady-state orbit is different from zero. Each time, a first simulation is started at $s=10^{-4}$ and a second one at $s=0.9$. Both are stopped when steady-state is reached. It should be noted that the vortex is not at equilibrium at the start of the simulations, and that the data acquisition takes place while the vortex core relaxes towards its steady-state position, determined by the value of the imposed current. This allows to capture the stiffness over the entire $s$-range, for any $J$. In addition, three different situations are investigated (see Fig.~\ref{fig:configs}), depending on the relative orientation between the in-plane curling magnetization and the AOF. A first set of simulations is performed without considering $\HOe$, then a set with $\HOe$ parallel to the curling magnetization, {\it i.e.}, $C=+1$, and finally a set with $\HOe$ antiparallel to the curling magnetization, {\it i.e.}, $C=-1$. Those are labelled below in this manuscript as noOe, $\Cp$ and $\Cm$, respectively. To evaluate the Amp\`ere-Oersted field, we consider an idealized situation where the current density is uniform in the pillar. Furthermore, as the thickness of the free layer is considerably smaller than the overall thickness of the pillar, edge effects are neglected, {\it i.e.}, an infinite cylinder is assumed (as in ref.\cite{araujo2022ampere}). The vortex core position $s$ as well as the energy components $\Wex$, $\Wms$ and $\WOe$ are internally computed by MuMax3 between each timestep.

The arbitrarily chosen functions for fitting  the energy components are even power polynomials of the 10$^\text{th}$ order, preceded by a pre-factor that depends on material and geometrical parameters. They have the following form
\begin{align}
    \Wex &= - \pi h A \sum^5_{i=0} a_i^\text{ex} s^{2i} \label{eq:WexF},\\
    \Wms &= 4\Ms^2 h^2 R \sum^5_{i=0} a_i^\text{ms} s^{2i}\label{eq:WmsF},\\
    \WOe &= - \frac{\pi}{5} C J \Ms R^3 h \sum^5_{i=0} a_i^\text{Oe} s^{2i}\label{eq:WOeF},
\end{align}
where $a_i^\text{ex}$, $a_i^\text{ms}$ and $a_i^\text{Oe}$ are the coefficients to be fitted relative to each term of the polynomials. Nonlinear least square fits are performed for $s\in[5\cdot 10^{-4},0.7]$, the upper value being close to the limit of vortex stability.\cite{araujo2022ampere} Using an adapted version of Eq.~(\ref{eq:Fr}), {\it i.e.}, $k=(\partial W / \partial s)/(s R^2)$, one can easily derive the expressions of the spring-like stiffness parameters, given as
\begin{align}
    \kex &= -\frac{\pi h A}{R^2}\sum_{i=1}^5 (2i) a^\text{ex}_{i}  s^{2(i-1)},
    \label{eq:kexF}\\
    \kms &= \frac{4\Ms^2 h^2}{R}\sum_{i=1}^5 (2i) a^\text{ms}_{i}  s^{2(i-1)},
    \label{eq:kmsF}\\
    \kOe &=- \frac{\pi}{5} \Ms R h \sum_{i=1}^5 (2i) a^\text{Oe}_{i}  s^{2(i-1)}.
    \label{eq:kOeF}
\end{align}
Let us add that a pretreatment is used before fitting. It consists in denoising the raw micromagnetic results thanks to a wavelet transform technique.\cite{lee2019pywavelets} This allows to obtain more accurate fitting parameters by getting rid of MuMax3 internal computational uncertainties. For every fit, we obtained a relative standard error on the intercept $a_0$ (see Eqs.~(\ref{eq:WexF})-(\ref{eq:WOeF})) of $0.00\%$. For the quadratic coefficient $a_1$ (resp. biquadratic $a_2$), it is always below $0.02\%$ (resp. $2.25\%$, while being below $1\%$ for most fits). Concerning the higher order coefficients $a_3$, $a_4$ and $a_5$, the errors were often a bit more important, which can be easily understood. Indeed as $s<1$ by definition, these terms contribute only to a limited extent to the total energy value.

\section{Results \& Discussion}
\begin{figure*}[ht]
    \centering
    \includegraphics[scale=1]{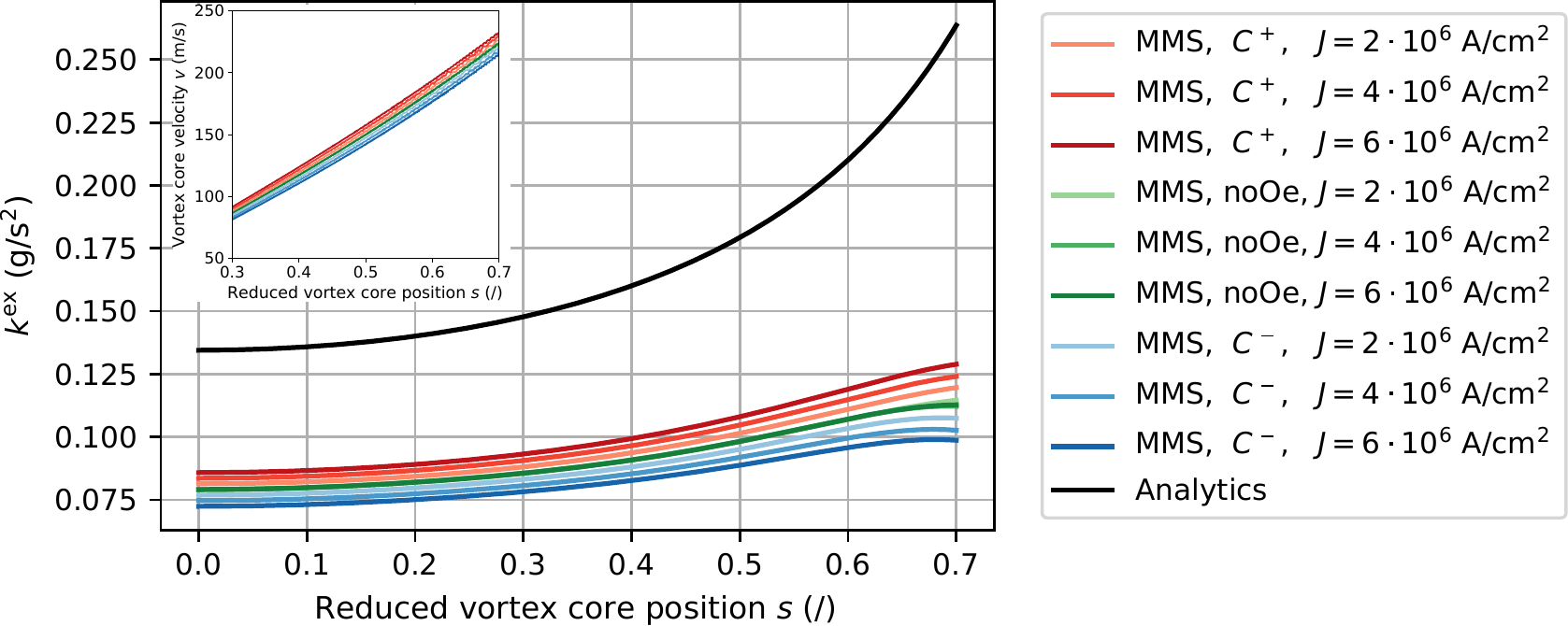}
    \caption{Exchange stiffness parameter $\kex$ as a function of the reduced vortex core position $s$. The analytical expression (black line, see Eq.~(\ref{eq:kexT})) is compared to micromagnetic simulation (MMS) results (colored lines, see Eq.~(\ref{eq:kexF})). The colors green, red and blue correspond to simulations without AOF (noOe), with AOF and $C=+1$ ($\Cp$) and with AOF and $C=-1$ ($\Cm$). The applied current densities $J$ were $2,~4$ and $6\cdot 10^6$~A/cm$^2$. MMS results were obtained after performing a nonlinear least squares fit on the exchange energy (see Eqs.~(\ref{eq:WexF})~\&~(\ref{eq:kexF})). The inset shows the vortex core velocity $v$ as a function of the reduced vortex core position $s$.}
    \label{fig:kex}
\end{figure*}

\begin{figure*}[ht]
    \centering
    \includegraphics[scale=1]{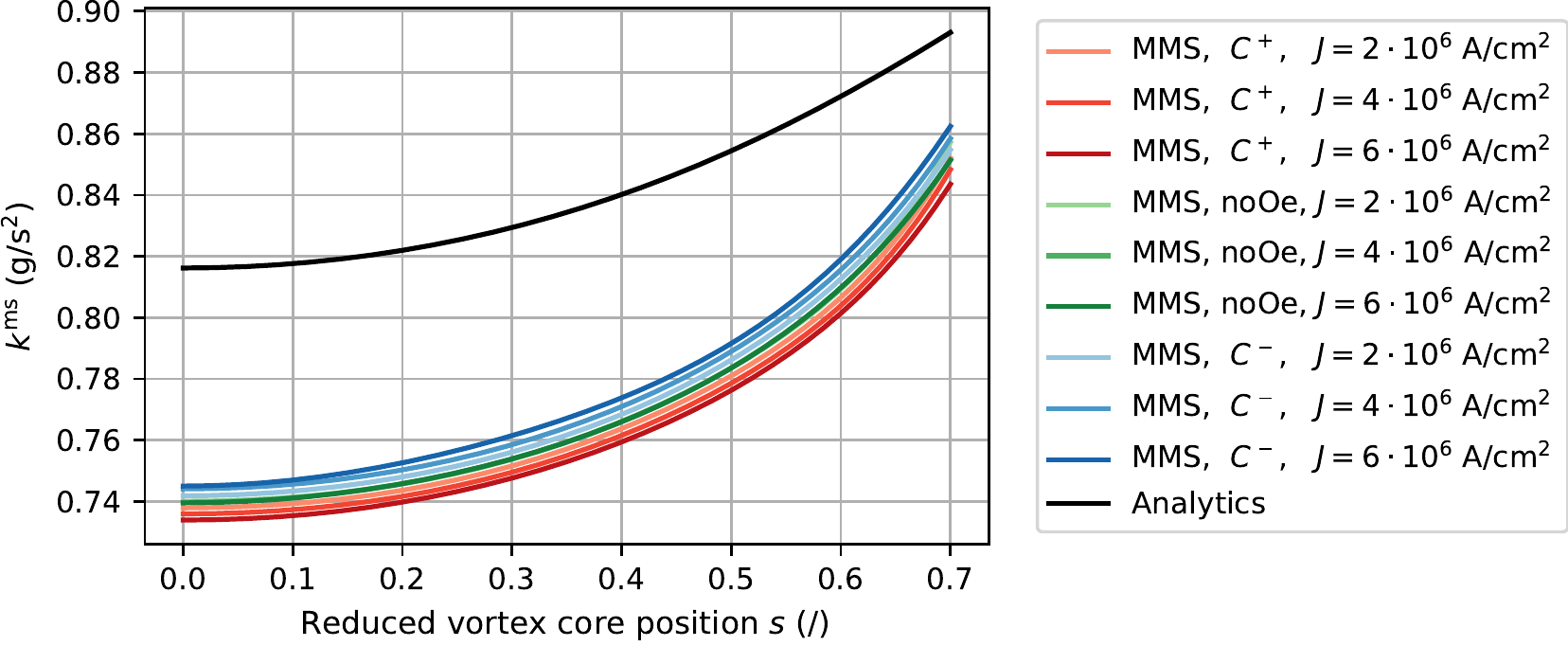}
    \caption{Magnetostatic stiffness parameter $\kms$ as a function of the reduced vortex core position $s$. The analytical expression (black line, see Eq.~(\ref{eq:kmsT})) is compared to micromagnetic simulation (MMS) results (colored lines, see Eq.~(\ref{eq:kmsF})). The colors green, red and blue correspond to simulations without AOF (noOe), with AOF and $C=+1$ ($\Cp$) and with AOF and $C=-1$ ($\Cm$). The applied current densities $J$ were $2,~4$ and $6\cdot 10^6$~A/cm$^2$. MMS results were obtained after performing a nonlinear least squares fit on the magnetostatic energy (see Eqs.~(\ref{eq:WmsF})~\&~(\ref{eq:kmsF})).}
    \label{fig:kms}
\end{figure*}

\begin{figure*}[ht]
    \centering
    \includegraphics[scale=1]{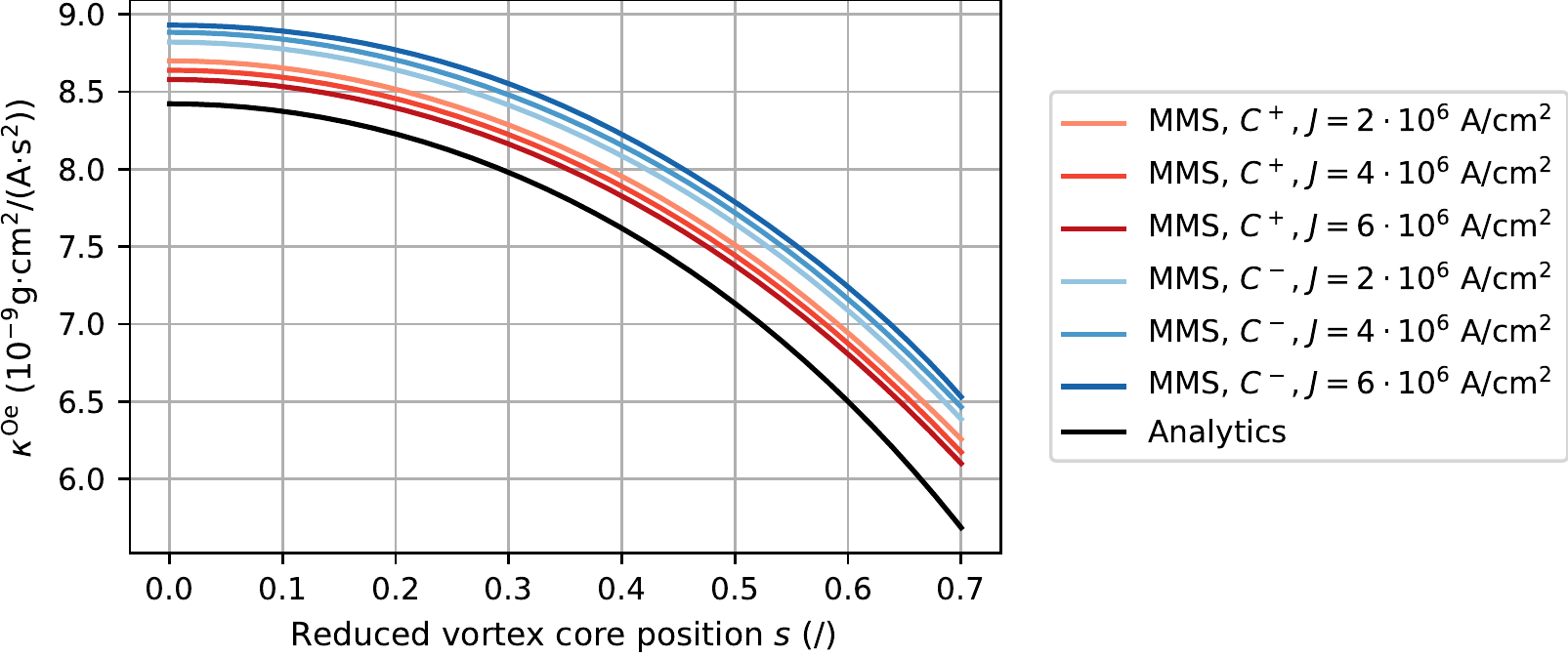}
    \caption{Amp\`ere-Oersted field (Zeeman-like) stiffness parameter $\kOe$ as a function of the reduced vortex core position $s$. The analytical expression (black line, see Eq.~(\ref{eq:kOeT})) is compared to micromagnetic simulation (MMS) results (colored lines, see Eq.~(\ref{eq:kOeF})). The colors red and blue correspond to simulations with AOF and $C=+1$ ($\Cp$) and with AOF and $C=-1$ ($\Cm$). The applied current densities $J$ were $2,~4$ and $6\cdot 10^6$~A/cm$^2$. MMS results were obtained after performing a nonlinear least squares fit on the Zeeman magnetic energy (see Eqs.~(\ref{eq:WOeF})~\&~(\ref{eq:kOeF})).}
    \label{fig:kOe}
\end{figure*}

Micromagnetic results (see Eqs.~(\ref{eq:kexF})-(\ref{eq:kOeF})) are compared to the analytical expressions presented previously (see Eqs.~(\ref{eq:kexT})-(\ref{eq:kOeT})). The evolution of the restoring parameters $\kex$, $\kms$ and $\kOe$ with respect to the vortex core position is depicted on Figs.~\ref{fig:kex}, \ref{fig:kms} \& \ref{fig:kOe}, respectively. Let us start with general observations valid for the three energy components. A slight disagreement is noticed between micromagnetism and TEA, even when the vortex core is at the center of the nanodot. These discrepancies could originate from two main hypotheses performed during TEA calculations. Firstly, the out-of-plane magnetization component $m_z$ is neglected for each derivation,\cite{gaididei2010magnetic, araujo2022ampere} due to the small area occupied by the vortex core. This simplification, necessary to obtain fully analytical expressions, is applicable as a preliminary approximation. However, even if the vortex core profile contribution is small, it is still different from zero. This is especially true for nanopillars of reduced radius, where the vortex core occupies a significant surface inside the free layer.  Secondly, the magnetization $\mathbf{m}$ is considered constant along the dot thickness. While appropriate for nanodots presenting a thickness of the order of the exchange length,\cite{metlov2002stability,guslienko2008dynamic,guslienko2006low} this assumption is not perfectly verified in micromagnetic simulations. In addition, supplementary deviations arise for an off-centered moving vortex core, explaining partly the TEA imperfect modeling when $s$ is evolving. Indeed within TVA, the vortex core is simply translated to a greater orbit while keeping its original shape. Such rigid motion assumption is justified by the fact that this ansatz was originally designed to study small displacements of static vortex cores. However, various groups\cite{khvalkovskiy2009vortex,fried2016localized,buchanan2006magnetic} have already shown the limitations of such theoretical framework to model accurately dynamic vortices. It has also been shown that a vortex core approaching the dot edge sees a growing trail of opposite magnetization appearing next to it.\cite{kasai2006current,gaididei2010magnetic, guslienko2008dynamic} Such dip, later responsible for the nucleation of a vortex of reversed polarity,\cite{hertel2007ultrafast} induces energy changes\cite{gliga2011energy} not perceived within TEA. 

Furthermore, a current induced splitting of the stiffness parameters can be observed, as already reported by Choi {\it et al.}. \cite{choi2008understanding} For increasing currents, the $\Cm$ and $\Cp$ curves move further apart from the noOe case. This behavior is not predicted at all by TEA calculations. These findings result from a modification of the spin distribution $\mathbf{m}$ caused by the AOF. Spins are tilted under its influence, leading them to deviate from the theoretical TVA magnetization. These effects will be discussed later in the manuscript. This distortion increases near the dot edges, considering the linear dependence of the AOF amplitude on the radial coordinate $r$ within the free layer. Moreover, the splitting increases with the current density as the AOF amplitude is proportional to the latter. This conclusion is supported by the fact that, in the noOe configuration, the stiffness is independent from the current value. Indeed, the mean difference in the simulations at $4$ and $6\cdot 10^6$~A/cm$^2$ compared with those at $2\cdot 10^6$~A/cm$^2$ is less than 0.17\% for the exchange stiffness, and less than 0.06\% for the magnetostatic stiffness (see green curves on Figs.~\ref{fig:kex} \& \ref{fig:kms}). A final point of interest is to note that for the exchange contribution, the favored configuration ({\it i.e.}, $C^+$) results in a stiffer force parameter, while the opposite is noticed for the other terms. As for now, we lack a robust justification for this phenomenon and further investigations would be required. After these preliminary observations, let us now look at each of the confinement contributions in more detail.

The TEA exchange stiffness parameter $\kex$ is the term presenting the most imprecise description compared to the simulated behavior (see Fig.~\ref{fig:kex}). For each $s$, the analytical value overestimates all MMS results. The predicted evolution is satisfying for a slightly off-centered vortex but diverges\cite{guslienko2001field,gaididei2010magnetic} to infinity for $s\rightarrow 1$, as expected from Eq.~(\ref{eq:kexT}) and the disregard of the vortex core profile in the analytical derivation. This does not reflect the behavior extracted from simulations as a maximum seems to appear at $s\approx 0.7$ (at least for $\Cm$ and noOe), followed by a decrease of the $\kex$ value. Various reasons could explain this drop. The most probable ones would be the interaction of the core with the dot edge and the 6-neighbor small angle approximation used by micromagnetic solvers, preventing accurate computation for large excitations. Moreover, a very large impact of the AOF is noticed. For a centered vortex core and $J=6\cdot 10^6$~A/cm$^2$, the exchange stiffness in $\Cp$ configuration is 18\% larger than the curve of opposite chirality. This value grows up to 31\% at $s=0.7$. Such considerable influence is linked to the fact that the gradient of magnetization $\nabla \mathbf{m}$ appears in the exchange energy formula, as shown in Eq.~(\ref{eq:W}). Any slight deviation in spin directions appearing in simulations with AOF has thus a major impact on the stiffness results compared to calculations using the unaffected theoretical distribution. Besides, the rotational velocity $v$ of the vortex core can give an indication of its degree of deformation (see inset in Fig.~\ref{fig:kex}), especially when compared with the critical velocity $v_\text{cr}\simeq 1.66 \gamma \sqrt{A}$ where $\gamma$ is the gyromagnetic ratio.\cite{lee2008universal} The velocities reached at the largest oscillation orbits are gradually approaching $v_\text{cr}=302$~m/s, suggesting a large dip following the vortex core. An impact of the chirality on the velocity was to be expected, given the frequency splitting reported in ref.\cite{araujo2022ampere}. However, the impact of the current magnitude on the velocity, for the same chirality and vortex core position, shows its influence on the magnetization dynamics.

In Figure~\ref{fig:kms}, one can observe that, as expected, the magnetostatic confinement dominates largely both other terms,\cite{guslienko2006magnetic,guslienko2006low} at least for the range of currents explored. For dots with wider radii, or at very large current densities, the AOF stiffness parameter $k^\text{Oe}$ should gain relative importance. The analytical expression of $\kms$ overestimates the value extracted from simulations (for the noOe case) by 10\% at $s=0$. This result is consistent with the overestimation of the first critical current $J_\text{c1}$ and eigenfrequency $\omega_0$ observed in our previous study\cite{araujo2022ampere} when using TEA, as $J_\text{c1}$ and $\omega_0 \propto \kms(s=0)$. As depicted in Fig.~\ref{fig:kms}, the magnetostatic confinement constantly increases as the core moves towards the edge, for both methods. Though the growth is limited, as $\kms$ is only 15\% greater at $s=0.7$ compared to the centered vortex case (for noOe). The analytical function models well the simulated evolution of $\kms$ with respect to $s$ for low vortex displacements. For greater $s$ values however, discrepancies are observed between the two methods. This difference is partially explained by the difficulty to obtain an analytical expression for the stray field.\cite{gaididei2010magnetic} The AOF seems to have a limited impact on the magnetostatic energy, all curves being close to the noOe configuration.

Furthermore, the analytical expression of $\kOe$ underestimates simulation results (see Fig.~\ref{fig:kOe}), for both $\Cp$ and $\Cm$. However the predicted behavior describes finely its evolution on the whole $s$ range. Such accuracy was bound to occur given the easier derivation of the Zeeman energy,\cite{araujo2022ampere} compared to non-local magnetostatic interactions. It may also be due to the higher order polynomial used in Eq.~(\ref{eq:kOeT}). More fundamentally, the AOF does not present any out-of-plane component as it appears in a plane perpendicular to the current direction, if edge effects are neglected.\cite{araujo2022ampere} Following Eq.~(\ref{eq:W}), the out-of-plane vortex magnetization $m_z$ has thus a negligible impact on the Zeeman energy. Nevertheless, constantly decreasing curves are obtained, $\kOe$ being 40\% greater for a centered vortex than in $s=0.7$. In addition, a non-negligible impact of the AOF is again perceived on its own stiffness parameters $\kOe$, in the simulations. This result was expected, as we get a modification of the magnetization distribution $\mathbf{M}$ appearing in Eq.~(\ref{eq:W}). The splitting stays roughly constant in amplitude, irrespective of the vortex core position.

\begin{figure*}[ht]
    \centering
    \includegraphics[scale=1]{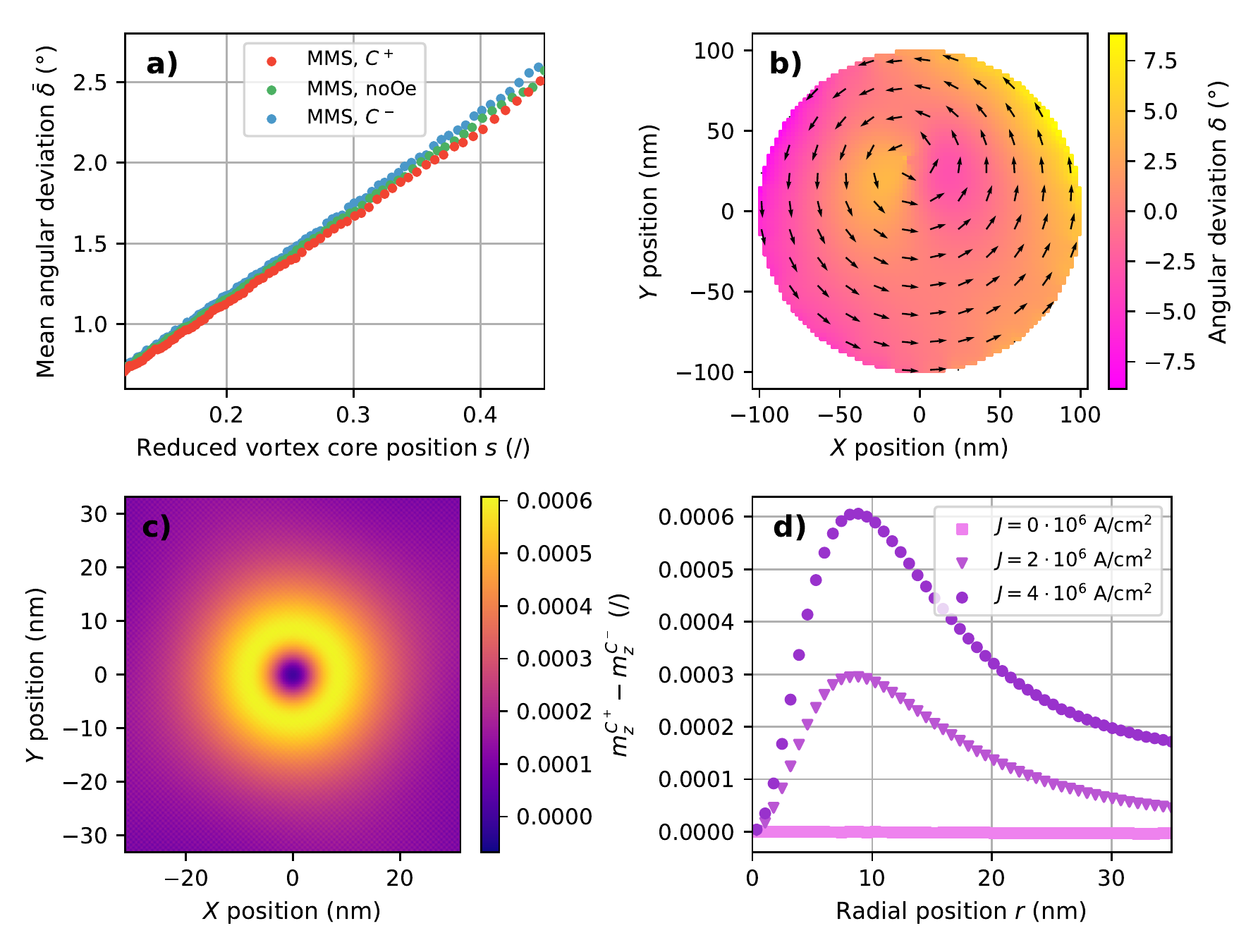}
    \caption{Impacts of the Amp\`ere-Oersted field on the magnetic texture ($J=4\cdot10^6$~A/cm$^2$, if not specified). (a) Mean angular deviation between the planar magnetization vector predicted by the two vortex ansatz and the one retrieved from simulations as a function of the reduced vortex core position. The colors green, red and blue correspond to simulations without AOF (noOe), with AOF and $C=+1$ ($\Cp$) and with AOF and $C=-1$ ($\Cm$). (b) Out-of-plane snapshot of the angular deviation for a dynamic off-centered vortex core, in each cell of the free layer. (c) Comparison between the out-of-plane magnetization components of two centered vortices of opposed chiralities, under the same current. (d) Cross-section of the difference in $m_z$ profiles at the center of the disk, at various current densities $J$.}
    \label{fig:texture}
\end{figure*}

Finally, let us look at the magnetic texture and the influence of the AOF on it. In Figure~\ref{fig:texture}a we define the mean angular deviation as $\bar{\delta}=\sum^N_i |\delta_i|/N$, where $\delta_i$ is the angle between the planar magnetization vector $(m_x,m_y)$ predicted by the TVA (see Eqs.~(\ref{eq:angles})~\&~(\ref{eq:m})) and the one retrieved from simulations, for a given micromagnetic cell, and $N$ is the total number of cells. The effects of the core deformation will be discussed afterwards. The mean deviation $\bar{\delta}$ is close to zero for low $s$ values and increases the further the vortex core is located from the center, as expected given that the TVA was designed for statics. What is more useful here is to use the TVA as a reference to compare the three configurations we are examining. As already suggested by the inset in Figure~\ref{fig:kex}, one can see that the deformation of the magnetic texture depends on the relative orientation between the in-plane swirling spins and the AOF. In addition, the splitting of the angular deviation widens as the core approaches the edge of the magnetic disc. To better understand these results, an out-of-plane snapshot of the angular deviation for each cell of the disc during a simulation is available in Figure~\ref{fig:texture}b. As anticipated, the regions that deviate most from the TVA predictions are those close to the core ({\it i.e.}, the dip) and at the edges of the disc. As the core moves towards the center, the deviations gradually become smaller.

In Figure~\ref{fig:texture}c, one can observe the deformation of the core profile for a centered vortex. To do so, we compare the out-of-plane magnetization components of two free layers under the same current but containing vortices of opposite chiralities. A more stringent cell size of $0.5\times 0.5\times 2.5$~nm$^3$ was used for a finer precision. As $||\mathbf{m}||=1$, we plot $m_z^{C^+}-m_z^{C^-}$ to highlight any variation in the planar magnetization components. At exactly $(0,0)$, there is no difference between both situations as the amplitude of the AOF is zero at these coordinates. In every other points of the vortex core though, the magnetization is confined in the plane by the AOF, modifying the value of $m_x$ and $m_y$. By conservation of the norm, the value of $m_z$ changes by an amount that depends on the relative orientation between the planar magnetization and the magnetic field. Such effect is comparable to what was reported by Dussaux {\it et al.}\cite{dussaux2012field} for a perpendicular magnetic field, {\it i.e.}, $m_z = H_\text{perp}/H_\text{s}$ (with $H_s$ the saturation field), where part of the magnetization was imposed outside the vortex core by the addition of an external field. In our case however, little effect is visible outside of the vortex core. In fact, as the magnetization is planar for both chiralities in this region, $m_z$ (which is nearly zero) is not impacted by the confinement. For reference, the half width at half maximum of the vortex core profile is $\sim 7$ nm. In Figure~\ref{fig:texture}d, one can see that the maximum deformation depends linearly on the amplitude of the current. This effect is a direct evidence of the influence of $J$, and therefore of the AOF, on the magnetic texture. The choice of working at $s=0$ was made to ensure that the core is at exactly the same location regardless of the chirality, even if the deformation is fairly small at this position. One last important thing to note is that part of this distortion originates from the spin-transfer torque. However, the further the core is from the center, the greater the relative contribution of the field to the deformation, as its amplitude increases with the position. It should be added that the value of the energy at $s=0$ is of little importance for the stiffness parameters, since those are rather linked to the shape of the potential well.

To put the results presented in this work into perspective, we believe that our restoring parameter expressions could be integrated into existing TEA models, to replace analytical expressions. Following such a data-driven approach, {\it i.e.}, deriving $\kex$, $\kms$ and $\kOe$ from a limited set of micromagnetic simulations, one could obtain more reliable results from TEA compared to micromagnetism and, by extension, to experimental results. The downside being that these parameters would only be valid for a given geometry and material parameters. Moreover, to obtain expressions valid for a continuous range of currents, one should proceed to an interpolation between a few curves selected cleverly, as the splitting increases with $J$. Nevertheless, this constitutes an alternative solution for a more accurate analytical model describing STVO dynamics. 

Although the influence of the temperature was not taken into account in this study, one may still discuss its potential effect on the restoring forces. It is well known that a magnetic tunnel junction powered by large current densities sees its temperature rise,\cite{prejbeanu2004thermally,prejbeanu2013thermally,strelkov2018impact} by up to several tens of degrees.\cite{lee2008increase} This Joule heating is especially important near the tunnel barrier (typically made of MgO) as its resistance is orders of magnitude higher than the one of the metallic layers. More importantly, the temperature reached depends on the magnitude of the current.\cite{lee2008increase} This means that the splitting phenomenon of the restoring forces, only caused by the current induced Amp\`ere-Oersted field in the present study (as $T=0$~K), could be enhanced by the larger temperature increase for higher currents. Indeed, any local random fluctuation of the spin texture $\mathbf{m}$ due to thermal effects impacts directly the value of the energy components, following Eq.~(\ref{eq:W}). It is clear that these effects, as well as the impact of the temperature on the different material parameters, could be considered in future works to further improve the validity of TEA-based models.

\section{Conclusion}
Restoring forces appearing in off-centered magnetic vortex states were examined theoretically. Exchange, magnetostatic and Zeeman stiffness parameters were obtained from simulations and compared to analytical expressions from the literature. To do so, the energy components were directly extracted from MuMax3, then fitted to high-order polynomials. The derivatives of these functions allowed to calculate stiffness parameters, analogously to what is done for classical springs under deformation. Discrepancies between the Thiele equation approach results and micromagnetic simulations were observed for each term, such as shifts of the curves and disagreeing behavior for large relative core position value. These differences were expected given the assumptions used for the theoretical derivations. More importantly, a chirality dependent splitting was observed in the stiffness value, with a deviation depending on the input current intensity. We provide evidence that this phenomenon is the result of a modification of the spin distribution caused by the current induced Amp\`ere-Oersted field. This hypothesis is supported by the fact that the stiffness was independent of the current imposed when the AOF was not taken into account. Finally, we believe that the expressions we derived from simulations could be implemented into existing STVO models to better render experimental results.
\section*{Acknowledgements}
Computational resources have been provided by the Consortium des Équipements de Calcul Intensif (CÉCI), funded by the Fonds de la Recherche Scientifique de Belgique (F.R.S.-FNRS) under Grant No. 2.5020.11 and by the Walloon Region. F.A.A. is a Research Associate and S.d.W. is a FRIA grantee, both of the F.R.S.-FNRS.

\end{document}